\documentclass{aastex62}

\usepackage{amsmath}
\graphicspath{{./}{figures/}}
\usepackage{graphicx}

\submitjournal{ApJ}

\begin{document}

\title{Relativistic Astronomy. II. In-Flight Solution of Motion and Test of Special Relativity Light Aberration}

\author{Jin-ping Zhu}
\affil{Department of Astronomy, School of Physics, Peking University, Beijing 
100871, China; zhujp@pku.edu.cn}

\author{Bing Zhang}
\affil{Department of Physics and Astronomy, University of Nevada Las Vegas, NV 89154, USA; zhang@physics.unlv.edu}
\affiliation{Department of Astronomy, School of Physics, Peking University, Beijing 
100871, China; zhujp@pku.edu.cn}
\affiliation{National Astronomical Observatories, Chinese Academy of Sciences, Beijing 100012, China}

\author{Yuan-pei Yang}
\affil{Kavli Institute for Astronomy and Astrophysics, Peking University, Beijing 100871, China}
\affiliation{National Astronomical Observatories, Chinese Academy of Sciences, Beijing 100012, China}

\begin{abstract}
The Breakthrough Starshot project aims to send centimeter-sized, gram-scale ``StarChip" probes to Alpha Centauri at a speed of $\sim0.2c$. On the other hand, Zhang \& Li recently proposed that trans-relativistic cameras may be sent to any direction to study astronomical objects and test special relativity. To conduct such ``relativistic astronomy'', one needs to solve the motion of the probe in flight.
We solve the motion of the probe (including the moving direction and the velocity) by comparing the positions
of three or more point sources observed in the Earth rest frame and in the probe's comoving frame.
When the positions of enough point sources are taken into account, the motion of the probe can be solved with an error which is even smaller than the diffraction limit of the transrelativistic camera.
After solving the motion, when the measurement of the position of an additional point source is introduced, one can use the data to test the light aberration effect in special relativity.
The upper limit of the photon mass can be placed from the deviation of aberration to slightly lower than the energy of the photon, e.g. $\sim 1 \rm {eV}$.

\end{abstract}


\section{Introduction} \label{sec:intro}
The Breakthrough Starshot is a program of the Breakthrough Initiatives\footnote{https://breakthroughinitiatives.org}
with the aim of proving the concept of developing ultra-fast light-driven probes.
The centimeter-sized, gram-scale ``StarChip" probes, each carrying 4 sub-gram-scale 2 megapixel cameras, 
are expected to be accelerated to a speed of $\sim0.2c$, which is projected to reach Alpha Centauri in $\sim20$ 
yr from the launch and transfer back to earth the images of the exoplanet Proxima b orbiting Proxima Centauri 
$4.37\ \rm {ly}$ away \citep{2016Natur.536..437A}.

When a probe travels with a transrelativistic speed, some interesting relativistic effects can be observed. For example, \cite{christian17}
presented a method of measuring acceleration of a probe using the temporal Terrell effect \citep{terrell59,penrose59}. To perform such measurements, the probe needs to be close to astrophysical masses, so that the size of the objects can be measured.

\cite{zhangli18} (Paper 1) suggested that one can actually use relativistic cameras more generally. Instead of aiming them to a particular target (e.g. Alpha Centauri), they suggested that one can send these cameras to any direction to study the universe. As natural spectrographs, lenses, and wide-field cameras, relativistic cameras can be used to observe the universe in an unprecedented manner and to perform unique tests on relativity. \cite{zhangli18} termed this approach of studying the universe ``relativistic astronomy''.

With a transrelativistic camera in space, one important task is to solve the motion of the camera (including both the direction of motion and the speed of the camera), so that the Doppler factors in all the directions in the sky can be solved precisely. One may use the information of the initial laser acceleration configuration or the telecommunication signals to determine these parameters. For a long enough observation, one can use the relative motion of the distant objects, i.e. the optical flow, to infer the motion of the probe. However, after acceleration is finished, the camera will travel in the geodesic trajectory defined by the gravitational field of all the masses (e.g. Sun, Earth, Moon, and other large planets), so that both the direction and the speed of the camera are subject to change. The telecommunication method requires multiple receiving stations, and becomes progressively difficult as the probe travels far away from Earth. On the other hand, it is possible to use the in-flight ``snapshot'' image of some point sources and compare them with the image taken from Earth \citep{zhangli18}.
This paper addresses this problem in great detail and show that the motion of the camera can be indeed solved using the information of three or more point sources (Section \ref{sec:al}).

A transrelativistic camera also allows a direct test of the light aberration effect in an unprecedented regime \citep{zhangli18}. 
\cite{hirshfeld01} tested aberration of light by observing the parallaxes of distant stars. 
\cite{kopeikin07} measured a small gravitational aberration of light by Earth-based experiments. Due to the slow motion and weak gravity of the earth, the precision of these tests was limited. 
In this paper, we present a detailed treatment to perform more precise constraints on the deviation of the aberration angle in special relativity  and a constraint on the rest mass of the photon by the aberration measurements (Section \ref{sec:constraint}).


\section{Solving the motion of a transrelativistic camera} \label{sec:al}

\subsection{Method}\label{subsec:method}

When a camera travels in the interstellar space with a transrelativistic speed,
the positions of celestial objects will be more concentrated in the moving direction 
due to the relativistic effects. Suppose that a probe carrying a camera moves in a certain
direction with a constant speed. One can define two rest frames: 
the Earth rest frame $K$, and the probe's comoving
frame $K'$\footnote{More precisely, one should use the barycenter rest frame of the solar system rather than the Earth frame for the rest frame $K$. This will introduce an additional small correction factor scaled with the Earth motion speed ($\beta \sim 10^{-4}$), which is much smaller than the probe's motion speed. Also since we care about the snapshot images taken from both frames, the optical flow effect can be neglected. }. Let us define the Lorentz factor of the probe as $\Gamma = 1/\sqrt{1 - 
\beta^2}$, where $\beta = v / c$ is the normalized speed of the probe. 
The angle between the object moving direction and the line of sight 
in two different frames are related through the relation of light aberration \citep{rybicki79}:
\begin{equation}
\label{E1}
\cos\theta' = \frac{\cos\theta + \beta}{1 + \beta\cos\theta}.
\end{equation}

\begin{figure}[htbp]
\centering\includegraphics[width=0.7\textwidth,trim=50 100 50 50,clip]{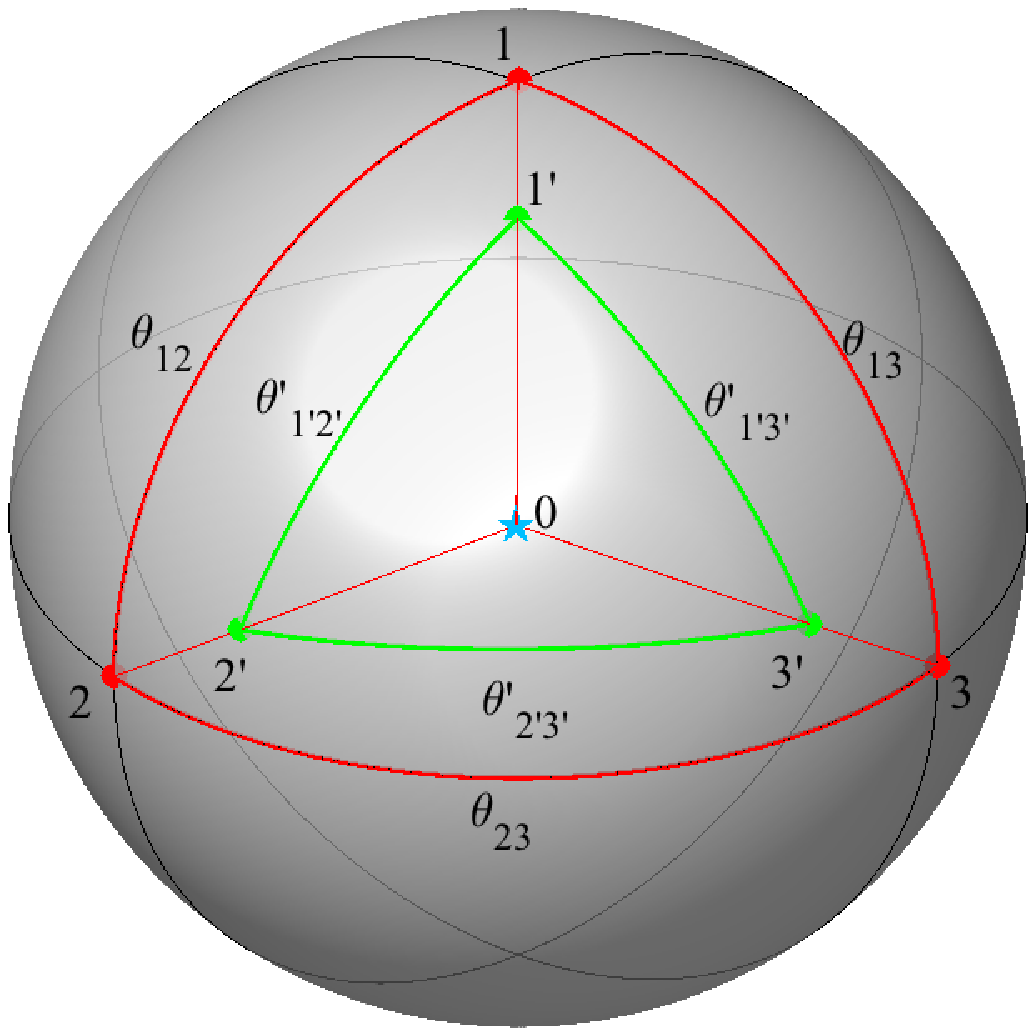}
\caption{Geometry to solve the motion of the probe. All the points and lines
are located on the celestial sphere. The three bright
points 1, 2, 3 in Frame $K$ are marked as red points, and their 
corresponding points 1$'$, 2$'$, 3$'$ in Frame $K'$ are marked as green
points. The blue star represents the direction of motion of the probe.
Here, the dimensionless velocity is set to $\beta = 0.5$.}
\label{fig:1}
\end{figure}

The parameters of the motion of the probe include the direction of motion and the dimensionless velocity $\beta$. 
In order to solve the motion of the probe, one needs to measure the sky positions of at lease three point sources in Frame $K'$ (1$'$, 2$'$, 3$'$) and compare them with the sky positions of the same three objects in Frame $K$ (1, 2, ,3) (see also \citealt{zhangli18}). 
Let us suppose that the direction of motion is to the point 0 and 0$'$ in Frame $K$ and
Frame $K'$, respectively. As seen from Figure \ref{fig:1}, the sky positions
of the sources in Frame $K'$ are located at the great circles defined by 
their corresponding sources in Frame $K$ and the direction of motion of the probe.
As a result, the opening angle of spherical triangle with the direction of motion as the vertex is unchanged after relativistic transformation, e.g. $\angle 102 = \angle 1'0'2'$.

We first consider a bottom-up (forward) approach of the problem assuming that the probe motion is known. 
The steps of solving the geometry of a three-point-source problem include the following:
\begin{enumerate}
\item Measure the parameters that define the direction of motion (i.e., right ascension $\alpha_0$ and
declination $\delta_0$ in the equatorial coordinate system) in Frame $K$ and the positions of the three point sources (i.e., $\alpha_i$, $\delta_i$, $i =1 , 2 , 3$).  
\item Use Eq.(\ref{A1}) (see Appendix \ref{app:GCD}) to obtain the angles 
between the direction of motion and three point sources in Frame $K$ 
(i.e., $\theta_{01}$, $\theta_{02}$ and $\theta_{03}$) as well as the angles
between three point sources themselves in Frame $K$ (i.e., $\theta_{12}$, $\theta_{13}$ and $\theta_{23}$).
\item Use Eq.(\ref{E1}) to obtain the angles between the direction of motion and the three point sources 
in Frame $K'$ (i.e., $\theta'_{0'1'}$, $\theta'_{0'2'}$ and $\theta'_{0'3'}$).
\item Use Eq.(\ref{A2}) (see Appendix \ref{app:ST}) to
obtain $\angle102$,  $\angle103$ and $\angle203$
in Frame $K$. 
The angles $\angle1'0'2'$, 
$\angle1'0'3'$ and $\angle2'0'3'$ in Frame $K'$ are 
equal to $\angle102$,  $\angle103$ and $\angle203$ in Frame $K$, respectively.
\item Use Eq.(\ref{A2}) (see Appendix \ref{app:ST}) to obtain the angles 
between the three point sources themselves in Frame $K'$ (i.e., $\theta'_{1'2'}$, 
$\theta'_{1'3'}$ and $\theta'_{2'3'}$). 
\end{enumerate}

Next, we consider a top-down (inverse) approach, in which we use the observed quantities to infer the information of motion, including the direction of motion (i.e., $\alpha_0$, $\delta_0$) and the the dimensionless velocity $\beta$. We start with three point sources as an example.
In practical observations, one can only obtain the positions of the three point sources with certain
observational uncertainties in both Frame $K$ (i.e., $\alpha_i$, $\delta_i$, $i =1 , 2 , 3$, and $\sigma_{\alpha_i}$, $\sigma_{\delta_i}$)
and Frame $K'$ (i.e., $\alpha'_i$, $\delta'_i$,  $i' =1 , 2 , 3$, and $\sigma_{\alpha'_{i'}}$, 
$\sigma_{\delta'_{i'}}$).
Using Eq.(\ref{A2}) (see Appendix \ref{app:ST}), one can obtain 
the corresponding angles in Frame $K'$ 
(i.e., $\theta'_{1'2'}$, $\theta'_{1'3'}$ and $\theta'_{2'3'}$).
The covariance error tensor $\pmb{\sigma}'_{\theta'}$ can be obtained by error propagation since $\theta'_{1'2'}$, $\theta'_{1'3'}$ and $\theta'_{2'3'}$ depend on $\alpha_1'$, $\delta_1'$, $\alpha_2'$, $\delta_2'$, $\alpha_3'$ and $\delta_3'$ in Frame $K'$.
One can also define an observational-value vector
$\pmb{\theta} = [\theta'_{1'2'} , \theta'_{1'3'} , \theta'_{2'3'}]^T$.
In order to calculate the position of the moving direction (i.e., $\alpha_0$, $\delta_0$)
and the dimensionless velocity $\beta$, one can try different model values based on the bottom-up (forward) approach and compare them against the data. The $\chi^2$ can be defined as
\begin{equation}
\label{E2}
\chi^2 = (\pmb{\theta}-\tilde{\pmb{\theta}})\pmb{\sigma}_{\theta'}'^{-1}(\pmb{\theta}-
\tilde{\pmb{\theta}}),
\end{equation}
where $\tilde{\pmb{\theta}} = [\tilde{\theta'}_{1'2'} , \tilde{\theta'}_{1'3'} , 
\tilde{\theta'}_{2'3'}]^T$ is the theoretical-value vector.
Going through a grid of the trial values of
the moving direction (i.e., $\alpha_0$, $\delta_0$) and the dimensionless velocity $\beta$ repeatedly, 
one can derive 
$\tilde{\theta'}_{1'2'}$, $\tilde{\theta'}_{1'3'}$ and $\tilde{\theta'}_{2'3'}$ for each trial, and 
use optimization algorithms, such as the Simulated Annealing algorithm \citep{metropolis53,kirkpatrick83,ingber89}, to find the minimum $\chi^2$ and identify the parameter optimum. 

After solving the motion with three point sources, one can apply error propagation to calculate the uncertainty of each parameter, i.e., $\sigma_{\alpha_0}, \sigma_{\delta_0}, \sigma_{\beta}$, and the correlation coefficient between each pair of the parameters, i.e., $\rho_{\alpha_{0}\delta_0}, \rho_{\alpha_{0}\beta}, \rho_{\delta_0\beta}$. 
The angles $\theta'_{1'2'}$, $\theta'_{1'3'}$ and $\theta'_{2'3'}$ depend on $\alpha_0 , \delta_0 , \alpha_1 , \delta_1 , \alpha_2 , \delta_2 , \alpha_3 , \delta_3, \beta$, which we assume that there is no correlation among the observational error of each parameter in Frame $K$. Then, the corresponding covariance matrix $\pmb{\sigma}_{\theta'}$ of $\theta'_{1'2'}$, $\theta'_{1'3'}$ and $\theta'_{2'3'}$ in Frame $K'$ can be obtained. On the other hand, we have obtained the covariance error tensor $\pmb{\sigma}'_{\theta'}$ when we calculate the parameter optimum. If $\pmb{\sigma}_{\theta'}$ and $\pmb{\sigma}'_{\theta'}$ are similar (i.e., the three eigenvalues of $\pmb{\sigma}_{\theta'}$ are equal to the three eigenvalues of $\pmb{\sigma}'_{\theta'}$, respectively), one can obtain the optimum of $\sigma_{\alpha_0}, \sigma_{\delta_0}, \sigma_\beta, \rho_{\alpha_{0}\delta_0}, \rho_{\alpha_{0}\beta}$ and $\rho_{\delta_0\beta}$.

\subsection{An example}

In the following, we use an example to show how one can use real data to solve the motion of a transrelativistic probe.  
Consider a transrelativistic camera on board a probe, which is moving towards the direction of Proxima Centauri and Proxima b.
One can estimate the position accuracy of the probe using the method described above.

We assume that the following parameters are independent variables, 
e.g., the dimensionless velocity $\beta$ of the probe, the angle $\theta_{\rm {c}}$ 
between the moving direction of the probe and the center of the camera's field of view in Frame $K'$, as well as 
the angular radius $r$ of the field of view of the camera on board the probe. Our goal is to use the measured data to constrain $\beta$ and $\theta_{\rm c}$ and see whether the input values can be reproduced. 

In Frame $K$, the datasets of the positions of bright celestial objects and their uncertainties can be directly obtained from the Data Release 2 \citep{gaia1} of the $Gaia$ mission \citep{gaia}.
More specifically, we select some sources with high precision 
($\sigma_\alpha, \sigma_\delta \lesssim 1\ \rm {mas}$) and high brightness ($G$-band 
magnitude $G \lesssim 4\ \rm {mag}$). The astrometric parameters for the $Gaia$ Data Release 2 sources that satisfy these criteria in the field of view of Proxima Centauri are presented in Table \ref{tab:ap}.
On the other hand, in Frame $K'$, the uncertainties of the celestial positions 
are approximately defined by the diffraction limit:
\begin{equation}
\label{E3}
\begin{split}
\theta\approx1.22\frac{\lambda}{D}=3.59\arcsec\ \left(
\frac{\lambda}{500\ {\rm nm}}\right)\left(\frac{D}{3 .5\ {\rm cm}}\right)^{-1},
\end{split}
\end{equation}
where $\theta$ is the angular resolution of the probe camera, $\lambda$ is the 
wavelength of light, and $D$ is the lens' aperture. 
With $\lambda\sim500\ \rm {nm}$ (the peak wavelength of sun-like spectrum)
and $D\sim3.5\ \rm {cm}$ (the size of the first prototype of Starshot cameras\footnote{
https://breakthroughinitiatives.org/News/12}), the
uncertainties of the positions in Frame $K'$ are $\sim3.59\arcsec$.

Table \ref{tab:E} shows some examples of solving the motion of the probe using observational data. The first four columns are the input parameters, including the probe dimensionless velocity $\beta$, the angular radius of the probe's field of view $r$, the angle between the moving direction and the center of the camera's field of view $\theta_{\rm c}$, and the aperture of the transrelativistic camera.  The fifth column lists the three HIP point sources chosen from the $Gaia$ catalog listed in Table \ref{tab:ap}. The three sources are chosen near the edge of the field of view, and we require nearly the same angular distance between each pair of the three points. Our top-down (reverse) approach can reproduce the input parameters $\beta$ and $\theta_{\rm c}$, with uncertainties $\sigma_{\alpha_0}$, $\sigma_{\delta_0}$, and $\sigma_{\beta}$ for the direction parameters $(\alpha_0, \delta_0)$ and the speed parameter $\beta$, respectively, listed in the next three columns. 

We start with the fiducial parameter set with $\beta = 0.2$, $r = 30\arcdeg$ and $\theta_{\rm c} = 0\arcdeg$.
The uncertainties of the moving direction are constrained to a precision of $\sigma_{\alpha_0}, \sigma_{\delta_0}\sim10^2\arcsec$ 
($\sim10^{-5}\ \rm {rad} $), and the uncertainty of the dimensionless velocity is $\sigma_\beta\sim10^{-5}$. Compared with the uncertainties of $Gaia$'s positions ($\sigma_\alpha, \sigma_\delta \lesssim 1\ \rm {mas}$) and the uncertainties of the camera in the probe (which is limited by the diffraction limit $\sim3.59\arcsec$), these uncertainties are larger (by a factor of $10^1-10^2$). We investigate how these uncertainties depend on the input parameters. We find that the dependences on $\beta$ and $\theta_{\rm c}$ are weak, while the dependence on $r$ is significant. Basically, if one shrinks the field of view, the position uncertainty would increase. We also test the dependence on the aperture of the camera $D$. The uncertainty inversely scales with $D$ as expected from the diffraction limit formula. We conclude that to achieve a better precision to solve the motion of the probe, cameras with a larger aperture $D$ and a larger field of view $r$ are preferred. The challenge of relativistic astronomy would be therefore trying to enlarge both $D$ and $r$ with the limited payload mass on the probe.


\subsection{Solving the motion with more than three point sources} \label{sec:2.3}

The position accuracy derived from three data points is typically much worse than the diffraction limit. We expect that the accuracy can be improved if one includes the information of more point sources. We further explore a method to solve the motion with $n\ge3$ point sources based on the general method described in Section \ref{subsec:method}. 

We use the {\em Monte Carlo} method to generate $n$ point sources with the assumption that all the point sources (i.e., $\alpha_i$ and $\delta_i$, $i = 1, 2, 3,\dots, n$) are located at the boundary of the probe's field of view. The fiducial parameter we set is $\beta = 0.2$, $r = 30\arcdeg$ and $\theta_{\rm c} = 0\arcdeg$. We assume that the uncertainties of the positions in Frame $K$ are $\sigma_{\alpha_i}, \sigma_{\delta_i}\sim1\ \rm {mas}$, while the uncertainties of the positions in Frame $K'$ are $\sigma_{\alpha'_i}, \sigma_{\delta'_i}\sim\ 3.59\arcsec$. 

When we estimate the position accuracy of the probe, a Markov Chain Monte Method (MCMC) based on {\tt emcee} package \citep{foremanmackey13} is adopted. The prior distributions of parameters are taken to be flat in the linear space. Since there are $n$ point sources to solve the motion, the dimensions of the observational-value vector $\pmb{\theta}$ and theoretical-value vector $\tilde{\pmb{\theta}}$  becomes $n(n-1)/2$. The covariance error tensor $\pmb{\sigma}'_{\theta'}$ becomes a $\frac{n(n-1)}{2}\times \frac{n(n-1)}{2}$ tensor. The objective function, i.e., log likelihood function, is therefore given by
\begin{equation}
\ln\mathcal L = -\frac{n(n-1)}{4}\ln2\pi-\frac{1}{2}\det\pmb{\sigma}'_{\theta'}-\frac{1}{2}(\pmb{\theta}-\tilde{\pmb{\theta}})\pmb{\sigma}_{\theta'}'^{-1}(\pmb{\theta}-
\tilde{\pmb{\theta}}).
\end{equation}

\begin{figure}
\plotone{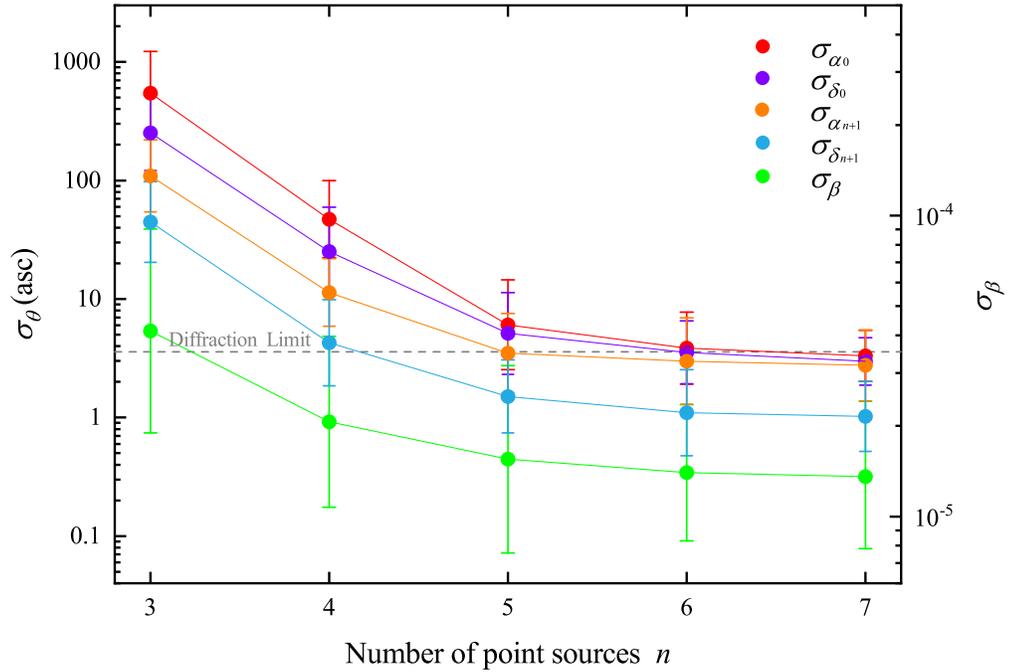}
\caption{The uncertainties of the moving direction, dimensionless velocity and position of the $(n + 1) ^ {\rm {th}}$ as a function of the number of selected sources used to solve the motion of the probe. The red, purple, orange, blue and green circle represent the uncertainties of $\sigma_{\alpha_0}, \sigma_{\delta_0}, \sigma_{\alpha_{n+1}}, \sigma_{\delta_{n+1}}$ and $\sigma_\beta$, respectively. All the uncertainties have a decreasing trend as a function of $n$, even though they converge to certain values as $n$ becomes greater than 5. The diffraction limit can be exceeded for  $\sigma_{\alpha_{n+1}}, \sigma_{\delta_{n+1}}$ and $\sigma_\beta$.}
\label{fig:2}
\end{figure}

We run each case 50 times. The results for the cases from 3 to 7 point sources are presented in Table \ref{table:3}. The uncertainties of the moving direction ($\sigma_{\alpha_0}$ and $\sigma_{\delta_0}$) and the dimensionless velocity ($\sigma_\beta$) have an obvious decrease trend (see Figure \ref{fig:2}) as a function of the selected point source number $n$. The decay rates of uncertainties will gradually decrease as the selected point source number $n$ increasing. The precision of $\sigma_{\alpha_0},\sigma_{\delta_0}$  can be even smaller than the camera's diffraction limit $\sim3.59\arcsec$ when $n>5$ is selected. We conclude that the direction of motion of the probe can be determined to the diffractive limit, if a large enough number of point sources (typically 5 or more) are selected to solve the motion.

\section{Tests of relativistic light aberration and constraints on the photon mass}\label{sec:constraint}

\subsection{Tests of relativistic light aberration effect}
With the motion of the probe solved, one can test special relativity light aberration formula by comparing the predicted position and the measured position \citep{zhangli18}. Selecting some point sources in Frame $K$, one can use the rotation transformation (see Appendix \ref{app:RT}) to calculate its celestial coordinate positions in Frame $K'$. For the cases of using three point sources to solve the motion of the camera, the results of position accuracy of the fourth point source are listed in the last four columns of Table \ref{tab:E}. The position uncertainties are generally $<60 \arcsec$. Similar to the constraints on the moving direction, the camera's field of view $r$ and aperture $D$ play the most important roles in determining the position precision. 
For different point sources with a similar angle $\theta_{04}$ between the fourth point and the moving direction in Frame $K$, the position uncertainties are similar.

If we use $n$ point sources to solve the motion of the camera, one can take the $(n+1)^{\rm th}$ point source to test the relativistic light aberration effect. We again use the {\em Monte Carlo} method to test how the precision of the test increases with $n$. After solving the motion, we randomly generate the $(n+1)^{\rm {th}}$ point source near the boundary of the probe's field of view.
The statistical results are presented in Table \ref{table:3}. The position uncertainties of the $(n+1)^{\rm {th}}$ point source also have a similar decrease relationship with the motion uncertainties as a function of $n$. Its position uncertainties are typically smaller than the diffraction limit $\sim3.59\arcsec$. 

\subsection{Constraints on the photon mass}

One specific mechanism to cause deviation of the aberration angle from the value predicted by special relativity is to invoke a non-zero mass of the photon 
\citep{debroglie22, debroglie23, debroglie40, proca36a, proca36b, proca36c, proca36d, proca37, proca38}. 
Many methods have been proposed to constrain the rest mass of the photon, e.g., 
the solar wind magnetic field \citep{ryutov97,ryutov07},
Coulomb's law \citep{williams71},
low frequency electromagnetic wave detection \citep{schumann52},
the frequency dependence of the speed of light \citep{lovell64,wu16,shao17},
and pulsar spin-down \citep{yang17}.
In the following, we constrain the photon mass using the limit of light aberration deviation measured from a transrelativistic camera.

If the photon has a non-zero rest mass $m_{\gamma}$, the Lorenz-invariant
dispersion is
\begin{equation}
\label{E5}
E=h\nu=\sqrt{p^2c^2+m_\gamma^2c^4}.
\end{equation}
The group velocity for a photon is
\begin{equation}
\label{E6}
v_{\rm g} = \frac{\partial E}{\partial p}=c\sqrt{1 - \left (\frac{m_\gamma c^
2}{h\nu}\right )^2}\simeq c\left[1 - \frac{1}{2} \left (\frac{m_\gamma c^
2}{h\nu}\right )^2 \right],
\end{equation}
where the last derivation holds when $m_{\gamma} \ll h\nu/c^2\approx2.48\ \rm eV$ for optical light.
Then, the aberration formula \citep{rybicki79} becomes
\begin{equation}
\label{E7}
\tan\theta'=\frac{v_{\rm g}\sin\theta}{\Gamma(v_{\rm g}\cos\theta+\beta c)},
\end{equation}
\begin{equation}
\label{E8}
\cos\theta'=\frac{v_{\rm g}\cos\theta  + \beta c}{\sqrt{(v_{\rm g} + \beta c\cos\theta)^2+\beta^2\sin^2\theta(c^2 - v_{\rm g}^2)}},
\end{equation}
For the case of $v_{\rm g} = c$ (i.e., $m_\gamma = 0$), Eq.(\ref{E8}) becomes
the light aberration formula in special relativity,  i.e., Eq.(\ref{E1}).

Suppose that we can obtain two images of the same sky area independently photographed by an Earth-based telescope in Frame $K$ and by a transrelativitic camera in Frame $K'$. Let us consider $n+1$ point sources that are well measured in both $K$ and $K'$. 
One can select $n$ of the $n+1$ point sources to solve the parameters of motion under the assumption of zero mass of the photon based on the method discussed in Section \ref{sec:al}. In theory, the position of the $(n + 1)^{\rm {th}}$ point source in Frame $K'$ can be calculated. Therefore, one can obtain a theoretical image ``photographed" by the transrelativitic camera. Matching the $n$ point sources in the observed image that are used to solve the motion with the $n$ corresponding point sources in the theoretical image, one can compare the position of the $(n + 1)^{\rm {th}}$ point source between observed image and the theoretical image in Frame $K'$. If the photon mass is non-zero, 
one would expect a slight deviation of the $(n + 1)^{\rm {th}}$ point source in two images due to the slight difference in the formulae of aberration of light. This is because the solution of motion and the angle between the direction of motion and $(n + 1)^{\rm {th}}$ point in Frame $K'$ would have been slightly different from the zero-photon-mass case. Using the upper limit on the mismatch between the $(n + 1)^{\rm {th}}$ point source of the two images, one can set an upper limit on the mass of the photon. 

We directly use the results of Section \ref{sec:2.3} to constrain the photon mass. Each run can obtain the uncertainties of the position of the $(n+1)^{\rm {th}}$ point source in Frame $K'$ under the assumption of zero mass of the photon. Next, we assume a non-zero photon mass and then solve the motion using the $m_\gamma$ aberration formula Eq.(\ref{E8}) to obtain the corresponding moving direction and dimensionless velocity. The position of the $(n + 1)^{\rm {th}}$ point source can be correspondingly calculated. By comparing this position with the case of non-zero photon mass, one can obtain the upper limit on the mass of the photon by requiring that the deviation between the two cases exceeds the position uncertainty of the $(n+1)^{\rm {th}}$ point source. With the number of selected point sources increasing to solve the motion, the limit on the photon mass is tighter. For the case of $n = 7$, the upper limit of photon mass is constrained to $\sim0.3\ \rm {eV}$. On the other hand, it won't become significantly lower than $1\ \rm {eV}$ defined by the energy of the optical band. This value is much greater than the maximum photon energy obtained using other methods \citep[e.g.][]{ryutov97,ryutov07,williams71,schumann52,lovell64,wu16,shao17,yang17}. Hence, optical relativistic imagery is not particularly useful in constraining the photon mass.

\section{conclusions} \label{sec:con}

In this paper, we present a detailed procedure to solve the motion of a transrelativitic camera using in-flight observational data. By comparing the positions of three sources observed in the Earth rest frame and in the probe's comoving frame, one can solve the motion of the probe including the direction and the velocity. We give some examples to show how one can use real data to solve the motion of the probe and estimate the precision of the motion with three point sources. By investigating the effect of the input parameters on the uncertainties of the motion, we conclude that a larger aperture and a larger field of view can give a better precision in solving the motion.

We further explore the approach to solve the motion with more than three point sources. By randomly generating point sources, we estimate the motion precision by an MCMC method. The precision increases with the increasing number of the selected point sources. The uncertainty of the moving direction can exceed the diffraction limit ($\sim3.59 \arcsec$), but would converge as $n$ is more than 5.  

With the motion of the probe solved, one can test the light aberration effect in special relativity. For the simplest case of using three point sources to solve the motion problem, the position uncertainty of fourth point source is typically smaller than that of the moving direction.
Again the camera's field of view and aperture play the most important roles in defining the position precision. We also investigate how $n$ point sources can improve the test.
 The uncertainty of the $(n + 1) ^ {\rm {th}}$ point source also decreases with an increasing $n$, which is typically smaller than the 
diffraction limit $\sim3.59\arcsec$.

The existence of the photon mass would introduce a slight difference in the formula of aberration of light. It causes a slight deviation of the position of $(n + 1) ^ {\rm {th}}$ point source compared to the zero $m_\gamma$ case. By requiring that the deviation of the position of $(n + 1) ^ {\rm {th}}$ point source between the $m_\gamma \neq 0$ and $m_\gamma = 0$ to be smaller than the uncertainty of the position of $(n + 1) ^ {\rm {th}}$ point source, one can place an upper limit on the photon mass. Our simulations show that the aberration method can constrain the photon mass to not too much below the photon energy $1\ \rm {eV}$ in the optical band. So this method is not competitive compared with other methods of constraining the photon mass.

\acknowledgments
The authors acknowledge anonymous referees for useful discussions. We also thank Rui Luo for helpful informations. This work is partially supported by the National Basic Research Program of China under grant No. 2014CB845800.
\software{Matlab, \url{https://www.mathworks.com}, Python, \url{https://www.python.org},  emcee \citep{foremanmackey13}}

\appendix

\section{Spherical Geometry}

Below we introduce some concepts in spherical geometry, the geometry of a two-dimensional spherical surface, that are used in this paper. 

\subsection{Great Circle Distance} \label{app:GCD}
Let us define ($\alpha_1$, $\delta_1$) and ($\alpha_2$, $\delta_2$) as the right ascension and declination in
radians of two points 1 and 2 on the sphere, respectively, and  $\Delta\alpha$ as the absolute difference in longitude.
The angle between the two points $\theta$ opened from the center of the sphere is defined by the spherical law of
cosines:
\begin{equation}
\label{A1}
\theta = \arccos(\sin\delta_1\sin\delta_2 + \cos\delta_1\cos\delta_2\cos(\Delta\alpha)).
\end{equation}
This can be straightforwardly derived when one of the poles is used as an auxiliary third point on the sphere.

\subsection{Spherical Triangle} \label{app:ST}

Suppose that three points (denoted as 1, 2, 3) on a sphere form a spherical triangle. 
The three angular distances among the three points ($\theta_{12}$, $\theta_{13}$ and $\theta_{23}$) and the measures of the three vertex angles 
of the triangle ($\angle132$, $\angle123$ and $\angle213$) are related by the cosine rule of the sphere triangle 
\begin{equation}
\begin{split}
\label{A2}
\cos\theta_{12} = \cos\theta_{13}\cos\theta_{23} + \sin\theta_{13}\sin\theta_{23}\cos\angle132,\\
\cos\theta_{13} = \cos\theta_{12}\cos\theta_{23} + \sin\theta_{12}\sin\theta_{23}\cos\angle123,\\
\cos\theta_{23} = \cos\theta_{12}\cos\theta_{13} + \sin\theta_{12}\sin\theta_{13}\cos\angle213.
\end{split}
\end{equation}

\section{Solving the parameters of positions in probe comoving Frame $K'$}\label{app:RT}

We adopt a representation of the coordinate transformation in terms of the
parameters of rotation --- the angle of rotation and the direction cosines
of the axis of rotation. The $rotation\ formula$ \citep{goldstein02} is
\begin{equation}
\label{D11}
\begin{split}
\textbf{r}'=\textbf{r}\cos\Phi+\hat{\textbf{n}}(\hat{\textbf{n}}\cdot\textbf{r})
(1-\cos\Phi)+(\hat{\textbf{n}}\times\textbf{r})\sin\Phi,
\end{split}
\end{equation}
where $\Phi$ is the rotation angle, $\hat{\textbf{n}}$ is the 
direction unit vector of axis, $\textbf{r}$ is the vector of the 
initial position, and $\textbf{r}'$ is the 
vector of the final position. 

If one knows the moving direction of the probe ($\alpha_0$ and $\delta_0$) and the
position of a random point ($\alpha_1$ and $\delta_1$) in the Earth rest Frame $K$, the 
corresponding position in the probe comoving Frame $K'$ ($\alpha'_1$ and $\delta'_1$) must be located at the great circle defined by the moving direction of the probe and the position of the random point in Frame $K$. In order to calculate the 
position in Frame $K'$ ($\alpha'_1$ and $\delta'_1$), one can rotate the 
cartesian coordinate position vector of the probe moving direction $\textbf{r}_0 = [\cos\delta_0\cos\alpha_0 , \cos\delta_0\sin\alpha_0 , 
\sin\delta_0]^T$ by an angle $\Phi = \theta'$ about the axis 
which is perpendicular to the plane of the great circle
and located at the center of the sphere, where $\theta'$ can be calculated from 
Eq. \ref{E1}. The axis is perpendicular to the plane of the 
great circle, so that the direction vector
of the axis is
\begin{equation}
\textbf{n}=
\begin{bmatrix}
n_x \\
n_y \\
n_z
\end{bmatrix}=\textbf{r}_0\times\textbf{r}_1=
\begin{bmatrix}
\cos\delta_0\cos\alpha_0 \\
\cos\delta_0\sin\alpha_0 \\
\sin\delta_0
\end{bmatrix}
\times
\begin{bmatrix}
\cos\delta_1\cos\alpha_1 \\
\cos\delta_1\sin\alpha_1 \\
\sin\delta_1
\end{bmatrix}=
\begin{bmatrix}
\cos\delta_0\sin\alpha_0\sin\delta_1 - \cos\delta_1\sin\alpha_1\sin\delta_0 \\
\cos\delta_1\cos\delta_1\sin\delta_0 - \cos\delta_0\cos\alpha_0\sin\delta_1 \\
\cos\delta_0\cos\alpha_0\cos\delta_1\sin\alpha_1 - \cos\delta_1\cos\alpha_1\cos\delta_0\sin\alpha_0
\end{bmatrix},
\end{equation}
where $\textbf{r}_1$ is the cartesian coordinate position vector of the random point
in Frame $K$. The direction unit vector is $\hat{\textbf{n}} = \textbf{n} / 
|\textbf{n}|$. As a result, the cartesian coordinate position vector of the corresponding
point $\textbf{r}'_1$ in Frame $K'$ can be calculated by Eq.(\ref{D11}). One
can then calculate the celestial position:
\begin{equation}
\alpha'_1 = \arctan\frac{r'_{1y}}{r'_{1x}},\delta'_1 = \arctan\frac{r'_{1z}}{\sqrt{r_{1x}'^2+r_{1y}'^2}}.
\end{equation}
Note that one should pay attention to the quadrant when calculating the celestial positions. One can finally obtain
$\sigma_{\alpha'_1}$, $\sigma_{\delta'_1}$ and the correlation coefficient
$\rho$ between them.

\begin{deluxetable*}{lccccc}[h!]
\tablecaption{Astrometric parameters from $Gaia$ DR2 \label{tab:ap}}
\tablecolumns{6}
\tablenum{1}
\tablewidth{0pt}
\tablehead{
\colhead{HIP Number\tablenotemark{a}} &
\colhead{$\alpha$\tablenotemark{b}/\arcdeg} &
\colhead{$\sigma_\alpha$\tablenotemark{c}/mas} &
\colhead{$\delta$\tablenotemark{d}/\arcdeg} &
\colhead{$\sigma_\delta$\tablenotemark{e}/mas} &
\colhead{$G$\tablenotemark{f}/mag}
}
\startdata
HIP 48002 & 146.775362659982 & 0.355484559318881  & -65.0719747969472 &
0.452647622691853 & 2.7888165 \\
HIP 50099 & 153.43379086538  & 0.381048026978585  & -70.0378756114751 &
0.412152102923925 & 3.1671705 \\
HIP 60260 & 185.338542066226 & 0.249153821072983  & -60.4007532768981 &
0.251904680424402 & 2.972596  \\
HIP 63003 & 193.648183007381 & 0.250793038754558  & -57.177982492016  &
0.287536909377994 & 3.9152856 \\
HIP 65109 & 200.147396031851 & 0.866965009254124  & -36.7126584300049 &
0.9151158414171   & 2.6858432 \\
HIP 68191 & 209.411620208564 & 0.110542550229435  & -63.6868421732935 &
0.113597426320103 & 4.332655  \\
HIP 68413 & 210.071883961754 & 0.0157461882226798 & -61.4811767464898 &
0.0212724498777446& 6.3933997 \\
HIP 70264 & 215.654157195748 & 0.606447518545649  & -58.4590342486566 &
0.645525706923748 & 4.505525  \\
HIP 70890\tablenotemark{g} & 217.393465742603 & 0.0576531066309365 & -62.6761821029238 & 0.104971236535274 & 8.953612  \\
HIP 71536 & 219.471584298892 & 0.435237848446469  & -49.4259545203336 &
0.43552937583903  & 4.1545763 \\
HIP 71908 & 220.624786167374 & 0.306439160958405  & -64.9761453374905 &
0.346400793582825 & 3.0168602 \\
HIP 72370 & 221.965348106629 & 0.353395987756381  & -79.0448239123319 &
0.249861855221442 & 3.2268114 \\
HIP 73036 & 223.892909779778 & 0.108083930105218  & -60.1146530932842 &
0.107148314251759 & 4.7807016 \\
HIP 73129 & 224.18320144613  & 0.29882749125488   & -62.7810549202754 &
0.28979783900068  & 6.3000674 \\
HIP 74376 & 227.983005446366 & 0.270611403896749  & -48.7380355294683 &
0.257024679570251 & 3.7678657 \\
HIP 75177 & 230.451054300018 & 0.4706870286551    & -36.2617424940659 &
0.430245831516788 & 2.804096  \\
HIP 75264 & 230.670180237484 & 0.627689424158833  & -44.6897202575442 &
0.523389157213469 & 3.1959176 \\
HIP 79882 & 244.580739905517 & 0.611283192789319  & -4.69233749011063 &
0.51437495211855  & 2.8429356 \\
HIP 80000 & 244.959020515699 & 0.408275536496331  & -50.1557296301489 &
0.325488297914412 & 3.6179643 \\
HIP 82363 & 252.446811482963 & 0.281259981504505  & -59.0414913798215 &
0.247374949109005 & 2.964356  \\
HIP 84405 & 258.835189375142 & 0.274003796302741  & -26.6077392836698 &
0.20226466988628  & 4.795791  \\
HIP 85258 & 261.324889804672 & 0.712821224324093  & -55.5299930306329 &
0.705065746762892 & 2.1833172 \\
HIP 86742 & 265.867953589588 & 0.632712365440421  & 4.56799013350183  &
0.6014778751505   & 2.2781103 \\
HIP 86929 & 266.433176038536 & 0.247410040661301  & -64.7241146561051 &
0.28347297854899  & 3.1453683 \\
HIP 92175 & 281.793608327785 & 0.379168248663183  & -4.74794370906114 &
0.357594134137972 & 3.7981637 \\
HIP 93085 & 284.432655909574 & 0.564336710740746  & -21.1067164143209 &
0.459084497805038 & 3.0392666 \\
HIP 95477 & 291.318920004013 & 0.463477890459631  & -24.5088046123265 &
0.426221125705944 & 4.8317966 \\
HIP 96229 & 293.523230152249 & 0.181975068418186  & 7.37826437436815  &
0.161013880989546 & 4.0115433 \\
HIP 100345& 305.252975453738 & 0.435907604437754  & -14.7814042849006 & 
0.317730995129632 & 2.7116423 \\
HIP 102488& 311.55470104855  & 0.700139915223491  & 33.9716578564345  &
0.812911932858741 & 2.0226774 \\
HIP 107315& 326.046610958014 & 0.546274241416486  & 9.87501784822262  &
0.600684192163677 & 1.7076352 
\enddata
\tablenotetext{\rm a}{The star mark from $Hipparcos$ catalogue \citep{perryman97}}
\tablenotetext{\rm b}{Right ascension at epoch J2015.0}
\tablenotetext{\rm c}{Standard uncertainties of right ascension}
\tablenotetext{\rm d}{Declination at epoch J2015.0}
\tablenotetext{\rm e}{Standard uncertainties of declination}
\tablenotetext{\rm f}{Magnitude in $Gaia$'s unfiltered band}
\tablenotetext{\rm g}{Proxima}
\end{deluxetable*}

\begin{deluxetable*}{cccccccccccc}[h!]
\tablecaption{Examples of error analysis \label{tab:E}}
\tablecolumns{12}
\tablenum{2}
\tablewidth{0pt}
\tablehead{
\colhead{$\beta$\tablenotemark{a}} &
\colhead{$r$\tablenotemark{b}/\rm {\arcdeg}} &
\colhead{$\theta_{\rm c}$\tablenotemark{c}/\rm {\arcdeg}} &
\colhead{$D$\tablenotemark{d}/\rm{cm}} &
\colhead{HIP Number\tablenotemark{e}} &
\colhead{$\sigma_{\alpha_0}$\tablenotemark{f}/$10^2$\arcsec} &
\colhead{$\sigma_{\delta_0}$\tablenotemark{g}/$10^2$\arcsec} &
\colhead{$\sigma_\beta$\tablenotemark{h}/$10^{-5}$} &
\colhead{HIP Number\tablenotemark{i}} &
\colhead{$\sigma_{\alpha'_4}$\tablenotemark{j}/$10^2$\arcsec} &
\colhead{$\sigma_{\delta'_4}$\tablenotemark{k}/$10^2$\arcsec} &
\colhead{$\theta_{04}$\tablenotemark{l}/\rm{\arcdeg}}
}
\startdata
 & & & & HIP 50099 & & & & HIP 48002 & 0.442 & 0.279\\
0.20 & $\sim$30 & $\sim$0  & 3.5 & HIP 65109 & 2.755 & 1.298 & 1.899 & HIP 75177 & 
0.331 & 0.207 & $\sim$25\\
 & & & & HIP 85258 & & & & HIP 86929 & 0.636 & 0.185\\
\hline
\hline
 & & & & HIP 50099 & & & & HIP 48002 & 0.413 & 0.258\\
0.15 & $\sim$30 & $\sim$0  & 3.5 & HIP 65109 & 3.345 & 1.599 & 1.877 & HIP 75177 &
0.303 & 0.194 & $\sim$25\\
 & & & & HIP 85258 & & & & HIP 86929 & 0.590 & 0.171\\
\hline
 & & & & HIP 50099 & & & & HIP 48002 & 0.387 & 0.239\\
0.10 & $\sim$30 & $\sim$0  & 3.5 & HIP 65109 & 4.564 & 2.216 & 1.849 & HIP 75177 &
0.277 & 0.182 & $\sim$25\\
 & & & & HIP 85258 & & & & HIP 86929 & 0.547 & 0.159\\
\hline
\hline
 & & & & HIP 63003 & & & & HIP 60260 & 1.538 & 0.648\\
0.20 & $\sim$15 & $\sim$0  & 3.5 & HIP 71536 & 8.183 & 3.859 & 4.250 & HIP 71536 &
1.110 & 0.701 & $\sim$15\\
 & & & & HIP 74376 & & & & HIP 82363 & 1.210 & 0.762\\
\hline
 & & & & HIP 68191 & & & & HIP 68413 & 26.730 & 12.019\\
0.20 & $\sim$5  & $\sim$0  & 3.5 & HIP 70264 & 152.837 & 64.293 & 18.645 & HIP 73036 &
26.530 & 11.601 & $\sim$5\\
 & & & & HIP 71908 & & & & HIP 73129 & 28.477 & 11.605\\
\hline
\hline
 & & & & HIP 75264 & & & & HIP 80000 & 0.342 & 0.189 & $\sim$20\\
0.20 & $\sim$30 & $\sim$45 & 3.5 &HIP 79882 & 2.051 & 1.090 & 5.729 & HIP 84405 &
0.261 & 0.191 & $\sim$45\\
 & & & & HIP 95477 & & & & HIP 92175 & 0.270 & 0.177 & $\sim$70\\
\hline
 & & & & HIP 86742 & & & & HIP 93085 & 0.253 & 0.206 & $\sim$60\\
0.20 & $\sim$30 & $\sim$90 & 3.5 & HIP 100345& 0.914 & 1.150 & 10.247 & HIP 96229 &
0.235 & 0.220 & $\sim$90\\
 & & & & HIP 107315& & & & HIP 102488& 0.264 & 0.191 & $\sim$120\\
\hline
\hline
 & & & & HIP 50099 & & & & HIP 48002 & 0.042 & 0.028\\
0.20 & $\sim$30 & $\sim$0  & 35 & HIP 65109 & 0.275 & 0.130 & 0.190 & HIP 75177 &
0.033 & 0.021 & $\sim$25\\
 & & & & HIP 85258 & & & & HIP 86929 & 0.064 & 0.019\\
\hline
 & & & & HIP 50099 & & & & HIP 48002 & 0.004 & 0.003\\
0.20 & $\sim$30 & $\sim$0  & 350 & HIP 65109 & 0.028 & 0.013 & 0.019 & HIP 75177 &
0.003 & 0.002 & $\sim$25\\
 & & & & HIP 85258 & & & & HIP 86929 & 0.006 & 0.002\\
\enddata
\tablenotetext{\rm a}{The dimensionless velocity}
\tablenotetext{\rm b}{Radius of the probe camera's field}
\tablenotetext{\rm c}{The angles between the moving direction of probe and centers of 
camera's field in Frame $K'$}
\tablenotetext{\rm d}{The lens' aperture of camera}
\tablenotetext{\rm e}{The star mark of three selected points from $Hipparcos$ catalogue \citep{perryman97}}
\tablenotetext{\rm f}{Standard uncertainties of right ascension of the moving direction of the probe}
\tablenotetext{\rm g}{Standard uncertainties of declination of the moving direction
of the probe}
\tablenotetext{\rm h}{Standard uncertainties of the dimensionless velocity}
\tablenotetext{\rm i}{The star mark of the forth points from $Hipparcos$ catalogue \citep{perryman97}}
\tablenotetext{\rm j}{Standard uncertainties of the fourth points' right ascension}
\tablenotetext{\rm k}{Standard uncertainties of the fourth points' declination}
\tablenotetext{\rm l}{The angles between the moving direction of the probe
and fourth points}
\end{deluxetable*}

\begin{deluxetable*}{lcccccc}[h!]
\tablecaption{The motion uncertainties when solving the motion with more than three point sources\label{table:3}}
\tablecolumns{7}
\tablenum{3}
\tablewidth{0pt}
\tablehead{
\colhead{$n$ \tablenotemark{a}} &
\colhead{$\sigma_{\alpha_0}$\tablenotemark{b}/$\arcsec$} &
\colhead{$\sigma_{\delta_0}$\tablenotemark{c}/$\arcsec$} &
\colhead{$\sigma_\beta$\tablenotemark{d}/$10^{-5}$} & 
\colhead{$\sigma_{\alpha'_{n + 1}}$\tablenotemark{e}/$\arcsec$} &
\colhead{$\sigma_{\delta'_{n + 1}}$\tablenotemark{f}/$\arcsec$}
}
\startdata
3 & 545.5$\pm$441.9 & 251.5$\pm$182.7 & 4.139$\pm$3.231 & 109.3$\pm$76.23 & 44.55$\pm$34.90 \\
4 & 47.16$\pm$35.19 & 25.19$\pm$21.58 & 2.064$\pm$1.350 & 11.33$\pm$7.429 & 4.283$\pm$3.583 \\
5 & 6.062$\pm$5.297 & 5.126$\pm$4.067 & 1.550$\pm$1.110 & 3.491$\pm$2.701 & 1.507$\pm$1.072 \\
6 & 3.855$\pm$2.691 & 3.543$\pm$2.161 & 1.399$\pm$0.730 & 2.993$\pm$2.509 & 1.100$\pm$0.919 \\
7 & 3.312$\pm$1.650 & 2.975$\pm$1.367 & 1.257$\pm$0.767 & 2.755$\pm$1.896 & 1.022$\pm$0.697 
\enddata
\tablenotetext{\rm a}{The number of selected points sources for solving the motion of the probe}
\tablenotetext{\rm b}{Standard uncertainties of right ascension of the moving direction of the probe}
\tablenotetext{\rm c}{Standard uncertainties of declination of the moving direction
of the probe}
\tablenotetext{\rm d}{Standard uncertainties of the dimensionless velocity}
\tablenotetext{\rm e}{Standard uncertainties of the $(n + 1) ^ {\rm {th}}$ points' right ascension}
\tablenotetext{\rm f}{Standard uncertainties of the $(n + 1) ^ {\rm {th}}$ points' declination}
\end{deluxetable*}

\end{document}